\documentclass[prl,11pt,reprint,nofootinbib,showpacs,showkeys]{revtex4-1}
\usepackage{comment}
\usepackage{natbib}
\usepackage[utf8]{inputenc} 
\usepackage{graphicx,color,overpic,mathtools}
\usepackage{amsthm,amsmath,amssymb,hyperref,mathrsfs}
\usepackage{braket,bm,bbm,setspace}
\PassOptionsToPackage{normalem}{ulem}
\usepackage{ulem} 
\usepackage{soul}
\usepackage{physics}
\usepackage{float}
\usepackage[makeroom]{cancel}
\usepackage[english]{babel}
\usepackage{graphicx}
\graphicspath{ {images/} }
\addto\captionsspanish{}
\usepackage[usenames,dvipsnames]{xcolor}
\hypersetup{colorlinks=true, citecolor=Plum, linkcolor=Plum,
urlcolor=Plum}

\begin{document}
\title{ On the nature of inner light-rings}
\author{Francesco Di Filippo}
\affiliation{Institute of Theoretical Physics, Faculty of Mathematics and Physics, Charles University, V Holešovičkách 2, 180 00 Prague 8, Czech Republic}

\begin{abstract}
Non-singular horizonless ultracompact objects provide a simple resolution to the black holes singularity problem. It has been shown that, if these objects are compact enough to exhibit the presence of the light-ring required to mimic the phenomenology of general relativity black holes, they must have at least one additional light-ring. The stability of the inner light ring has been proven under the assumption of Einstein equations and the validity of the null energy condition. Since this can have important repercussions on the instability of a horizonless ultracompact object and the existence of the latter requires some modified gravitational dynamics and/or exotic matter, it is desirable to obtain a model-independent proof of the stability of the additional light-ring.\\
In this paper, we prove the stability of the inner light-ring without any assumption on the dynamics of the theory, while assuming that the outer light-ring has the same properties as the Kerr light-ring.
Given the stringent observational constraints on the geometry at the outer light-ring scale, our result now rests solely on geometric considerations and applies to any theory of gravity with any matter content that cannot be ruled out by observations.
 
\end{abstract}
\maketitle
\section{Introduction}
Black holes represent both one of the biggest successes of general relativity and one of its long-standing challenges. While observational tests coming from the LIGO/Virgo 
\cite{PhysRevLett.116.061102,LIGOScientific:2019fpa,LIGOScientific:2020tif,LIGOScientific:2021sio}, the EHT \cite{EventHorizonTelescope:2019dse,EHT_Sgr}, and the GRAVITY \cite{GRAVITY:2018ofz} collaborations provide remarkable confirmations of the prediction of the theory, it is well known that general relativity also predicts its own limitations. A series of fundamental results teach us that physically realistic initial conditions will unavoidably produce a singular black hole spacetime \cite{Penrose:1964wq,Hawking:1970zqf,Senovilla:1998oua}. However, a singularity is nothing but a region missing from the spacetime that cannot be described by general relativity. Therefore, at least close to the singularity, we need to introduce new physics and we expect that singularities will not be produced once these effects are taken into account.

Among the possible classes of non-singular spacetimes \cite{Carballo-Rubio:2019fnb,Pandora,Carballo-Rubio:2021ayp} that can mimic the phenomenology of black holes, horizonless UltraCompact Objects (UCOs) are characterized by the absence of any horizon \cite{Mazur:2001fv,Mazur:2004fk,Cattoen:2005he,Barcelo:2009tpa,Carballo-Rubio:2017tlh,Arrechea:2021xkp,Jampolski:2023xwh} which makes it possible for deviations from to black hole spacetimes to be in principle observable. In particular, while the geometry is weakly constrained at the horizon scale, tests at the light-ring (LR) scale are very precise~\cite{Chirenti:2007mk,Vincent:2015xta,Cardoso:2017cfl,Carballo-Rubio:2018jzw,Cardoso:2019rvt,Volkel:2020xlc,Eichhorn:2022oma,Volkel:2022khh,Carballo-Rubio:2022imz,Carballo-Rubio:2023fjj}. Therefore, UCOs do not need to have an horizon, but must possess a LR which closely approximates the LR of Kerr black holes in order to pass the observational requirements.

In Ref.~\cite{Cunha:2017qtt} it was proved that an axisymmetric and stationary UCO that is compact enough to have at least one LR must contain at least one extra LR. Furthermore, the nature of the light-rings is not arbitrary.
Ref.~\cite{Cunha:2017qtt} also proves that at least one LR must be stable. This result is obtained either restricting to a spherically symmetric spacetime or with the use of the null convergence condition (which in general relativity is equivalent to the null energy condition). 
This is a crucial result in the understanding of the physics of UCOs as there is evidence that the presence of a stable LR leads to a non-linear instability of the UCO due to the piling up of perturbation at the minimum of the effective potential \cite{Cunha:2022gde}. This was explicitly shown for some models of boson stars \cite{Cunha:2022gde}, but the slow decay of the linear perturbation hints that the instability might be universal \cite{Cardoso:2014sna,Keir:2014oka,Franzin:2023slm} (see, however \cite{Guo:2024cts}).

However, the assumptions on the dynamics of the theory are an important restriction on the applicability of the result in \cite{Cunha:2017qtt}. Gravitational collapse in general relativity leads to the formation of Kerr black holes. Therefore, the creation of UCOs requires either new physics or exotic matter. The effects of such modifications cannot be negligible at the scale of the inner LR as the Kerr geometry does not possess a stable LR. 
Furthermore, we know that the null energy condition can be easily violated by quantum fields~\cite{Flanagan:1996gw}.
Therefore, it is definitely preferable to avoid any assumption on the physics at the inner LR scale. 

In this paper, we show that the conditions on the dynamics of the theory can be substituted by geometrical assumptions at the scale of the standard LR. In other words, we show that sny axisymmetric UCO invariant under simultaneous reflections $(t,\phi)\rightarrow(-t,-\phi)$ that has a LR which mimics the LR of a Kerr black hole must possess at least one stable LR for each rotation sense.

The requirement that the LR ``mimics" the LR of  a Kerr black hole will be made precise later on.
This generalization vastly increases the regime of applicability of the result, making it valid for any theory of gravity regardless of the matter content being considered. Of course, the requirement that the outer\footnote{In spacetimes that are not spherically symmetric, it might be not possible to define an outer and inner LR.  In most physically relevant scenarios this distinction should be very clear. If that is not the case, the ``outer" LR is, by definition, the one that mimics the properties of LRs in Kerr spacetime.} LR must mimic the property of a Kerr black hole can rule out some specific modifications of gravity. However, the main point is that all the assumptions are kinematical and at the scale of the outer LR. Therefore, a violation of the assumption of this investigation would lead to observational features.

\section{Properties of the effective potential}
Let us start by showing some properties of the effective potential that we are going to use later in the paper. We consider a UCO described by a stationary and axisymmetric solution. We consider the coordinates $(t,r,\theta,\phi)$, the symmetries of the geometry imply that $\partial_t$ and $\partial_\phi$ are Killing vectors, so all the metric components are functions of $r$ and $\theta$ only. If we solely consider spacetimes invariant under the transformation $(t,\phi)\rightarrow(-t,-\phi)$ to describe rotating objects around the axis of symmetry, the most generic line element we can write is (see e.g \cite{chandrasekhar_mathematical_1998})
\begin{equation}
ds^2=g_{tt}dt^2+g_{rr}dr^2+g_{\theta\theta}d\theta^2+2g_{t\phi}dtd\phi+g_{\phi\phi}d\phi^2\,.
\end{equation}
\\
The absence of horizon implies $g_{t\phi}^2-g_{tt}g_{\phi\phi}>0$. Furthermore, we require $g_{\phi\phi}>0$ to avoid the presence of closed timelike curves.
Light-rings are defined as regions in which particles moving along null geodesics only have momenta along $t$ and $\phi$. Their positions can be found studying the Hamiltonian for a free massless particle 
\begin{equation}\label{eq:Hamiltonian}
    \mathcal{H}=g^{\mu\nu} p_\mu p_\nu=g^{rr}p_r^2+g^{\theta\theta} p_\theta^2+V(r,\theta)=0\,.
\end{equation}
Where the potential $V$ is 
\begin{equation}
    V=\frac{1}{g^2_{t\phi}-g_{tt}g_{\phi\phi}}\left(
    E^2 g_{tt}+2E\Phi g_{t\phi}+\Phi^2 g_{\phi\phi}
    \right)\,.
\end{equation}
in which $E:=-p_t$ and $\Phi:=p_\phi$ are constants of motion. Eq.~\eqref{eq:Hamiltonian} and Hamilton's equations
\begin{equation}
    \dot{p}_\mu=-\partial_\mu\mathcal{H}=-\left(\partial_\mu\left(g^{rr}p_r^2\right)+\partial_\mu\left(g^{\theta\theta}p_\theta^2\right)+\partial_\mu V(r,\theta)\right)
\end{equation}
imply that the conditions for the light-rings $p_r=p_\theta=\dot{p}_\mu=0$ are equivalent to
\begin{equation}
  V(r,\theta)= 0\,,\qquad\text{and}\qquad  \partial_\mu V(r,\theta)= 0\,.
\end{equation}
Working directly with this potential is problematic as it also depends on the constants of motion of the test particle. However, it is possible to introduce two functions that only depends on geometrical quantities and whose extrema and saddle points coincide with the LRs but that \cite{Cunha:2017qtt}. To this end, we need to consider the potential functions\footnote{Note that this definition differs for a $\pm$ overall sign from the definition given in, e.g., \cite{Cunha:2017qtt}. This choice is not crucial, but it simplifies some signs later on. }
\begin{equation}\label{eq_H}
    H_\pm(r,\theta)=\pm\frac{-g_{t\phi}\pm\sqrt{g^2_{t\phi}-g_{tt}g_{\phi\phi}}}{g_{\phi\phi}}\,.
\end{equation}
The $+\,(-)$ subscript corresponds to the potential function for co(counter)-rotating geodesics. The absence of a trapping horizon implies the quantity in the square root is always positive. In spacetimes that contain a horizon, the effective potential would not be well defined within the trapped region. This is why the theorem does not hold if there is a trapping horizon (see also \cite{Cunha:2020azh,Ghosh:2021txu}).\\
 LRs correspond to the points for which the gradient of either one of the potential functions vanishes~\cite{Cunha:2016bjh}
\begin{equation}\label{eq:def_LR}
   \nabla H_+(r,\theta)=0\,\quad\text{or}\quad\nabla H_-(r,\theta)=0.
\end{equation}
Furthermore, at a light ring \cite{Cunha:2017qtt},
\begin{equation}
    \partial_\mu^2V=\frac{2\Phi^2}{\sqrt{g^2_{t\phi}-g_{tt}g_{\phi\phi}}}\partial_\mu^2H_\pm\,.
\end{equation}
Therefore, both the positions and the stability of the LRs can be studied by looking at the effective potentials $H_\pm$.

\textsl{\textbf{Boundary and asymptotic behavior.---}} \,
We now need to specify the asymptotic behavior of $H_\pm(r,\theta)$.
\begin{itemize}
    \item For very large values of the coordinate $r$, we need to recover asymptotic flatness for which $g_{tt}\rightarrow -1$, $g_{t\phi}\rightarrow 0$ and $g_{\phi\phi}\rightarrow r^2\sin^2{\theta}$. Leading to
    \begin{equation}\label{eq:bc1}
        H_\pm(r,\theta)\xrightarrow{r\rightarrow\infty}\frac{1}{r\sin{\theta}}\,.
    \end{equation}
    \item  Next, we study the behavior near the axis $\theta=0$ and $\theta=\pi$. In this limit, the absence of coordinate singularities implies that the ratio between the length $\mathcal{C}$ of a circumference around the axis and its proper radius must be equal to $2\pi$ in the limit of vanishing proper radius.  Denoting by $\rho$ the proper distance from the axis, this ratio is given by
    \begin{equation}
       2\pi= \lim_{\rho\rightarrow0} \frac{\mathcal{C}}{\rho}=\lim_{\rho\rightarrow0} \frac{1}{\rho}\int_{0}^{2\pi}\sqrt{g_{\phi\phi}}d\phi=2\pi\frac{\sqrt{g_{\phi\phi}}}{\rho}\,.
    \end{equation}
    Which implies
    \begin{equation}
        \lim_{\rho\rightarrow0}g_{\phi\phi}=\rho^2\,.
    \end{equation}
    Furthermore, to avoid curvature singularities, $g_{t\phi}$ must vanish at least as fast as $g_{\phi\phi}$ \cite{Cunha:2020azh}. We can now take the limit of the potential functions \eqref{eq_H}

\begin{equation}\label{eq:limit}
    \lim_{\theta\rightarrow 0}H_\pm(r,\theta)= \lim_{\theta\rightarrow \pi}H_\pm(r,\theta)=+\infty\,.
\end{equation}
    \item  Finally, we need to determine the behavior near the origin $r=0$. Because we are not considering wormhole-like objects\footnote{See \cite{Xavier:2024iwr} for a recent analysis of ultracompact traversable wormholes.}, the proper radius goes to zero near $r=0$. Proceeding with similar arguments to the ones that lead to Eq.~\eqref{eq:limit}, it is possible to prove that
    \begin{equation}\label{eq:bc3}
       \lim_{r\rightarrow 0} \mathcal{H_\pm}=+\infty\,.
    \end{equation}
\end{itemize}
Therefore, regularity conditions and asymptotic flatness fix the potential functions asymptotically and near the boundary of their domain of definition. This observation will be very relevant later.  

\section{Proof that LRs come in pairs}
Let us start by reviewing the analysis of Ref.~\cite{Cunha:2017qtt} which we will then generalize.\\
The proof of the theorem makes use of the Brouwer degree of a map (see e.g. \cite{dinca2021brouwer}) that can be defined as follows. Consider two compact, connected, and orientable manifolds $X$, $Y$ of equal dimension and
a smooth map $f : X \rightarrow Y$.
Let  $y_0 \in Y$ be a regular value such that the set 
$f^{-1}(y_0) = \{x_n\}$ 
has a finite number of points, with $x_n \in X$, such that $f(x_n) = y_0$, and the Jacobian $J_n = \det (\partial f /\partial x_n) \neq 0$. 
The Brouwer degree of $f$ with respect to $y_0$ can be defined as 
\begin{equation}
    \text{deg}f:=\sum_n\text{sign}\left(J_n\right)\,.
\end{equation}
To mention the crucial property of the Brouwer degree we need to recall that an \textit{homotopy} among function $f,g:X\rightarrow Y$ is a map 
\begin{equation}
    G:X\times [0,1]\rightarrow Y\,.
\end{equation}
such that
\begin{equation}
    G(x,0)=f(x)\,,\qquad G(x,1)=g(x)\,.
\end{equation}
The Brouwer degree of a map is invariant under homotopies for which $y_0\notin G\left( \partial X,t\right)$, i.e., invariant under continuous deformations for which there is never any boundary point whose image is equal to $y_0$.
In particular, this is always satisfied for transformations that do not change the map at the boundary (provided that $y_0$ is not in the image of the boundary for the map under consideration).

While all this discussion may seem very abstract, it becomes clear once we specify it for our setup. We apply the invariance of the Brouwer degree considering as map either one of the vectors fields $\textbf{v}_\pm$ with components 
\begin{equation}
    v_\pm^i=\partial^i H_\pm\,,\qquad i\in\{r,\theta\}\,.
\end{equation}
These vector fields map the 2-dimensional compact subset $X$ of the $\{r,\theta\}$ plane to a 2-dimensional space $Y$. The LRs are the points of $X$ which are mapped into the origin of $Y$. We choose $y_0$ to coincide with the origin of $Y$.  Therefore, the points $\{x_n\}$ correspond to the LRs. To ensure that the Brouwer degree is a topological invariant we need to show that at no point during the continuous transformation there is a LR at the boundary of $X$. This is very easy to do. In fact, Eqs.~\eqref{eq:bc1}, \eqref{eq:limit}, and \eqref{eq:bc3} show that the asymptotic values of the potentials functions and their derivatives are fully determined by regularity conditions and observational constraints. Therefore, it is enough to make $X$ large enough to ensure that the potential functions cannot have a LR at the boundary of $X$ and any allowed potential function can be continuously transformed into each other without changing the Brouwer degree.

The invariant Brouwer degrees of these maps are then
\begin{equation}
     \text{deg}\,v_\pm=\sum_n\text{sign}\left(\det \left(\partial_i \partial^jH_\pm(x_n)\right)\right)\,.
\end{equation}
So we can see that, for these specific maps, the determinant of the Jacobian is equal to the Hessian of the effective potential. Let us remind that the sign of the Hessian is positive for an extremum (either maximum or minimum) and negative for a saddle point. Since the Brouwer degree is zero for spacetimes that are not compact enough to admit a LR and is invariant under smooth deformations of the effective potential, it follows that the Brouwer degree must also be zero for any UCO admitting LRs. Therefore, the result of this theorem, which was proved in \cite{Cunha:2017qtt}, is that there is an equal number of saddle points and extrema of the effective potential. 

Ref.~\cite{Cunha:2017qtt} then shows that 
\begin{equation}
    G_{\mu\nu} p^\mu p^\nu = \partial_i\partial^i V\,,
\end{equation}
where $G_{\mu\nu}$ is the Einstein tensor and $p^\mu$ is a null vector field. If we assume general relativity and the null energy condition we obtain
\begin{equation}\label{eq:NEC}
    G_{\mu\nu} p^\mu p^\nu =T_{\mu\nu} p^\mu p^\nu > 0\,.
\end{equation}
Therefore, these assumptions on the dynamics of the gravitational theory and the matter content lead to the condition that the trace of the Hessian matrix must be positive. This implies that the extremum of the potential cannot be a maximum and it must be a minimum. 

We shall now prove the same result without any information regarding the dynamics of the theory.

\section{Kinematical proof that the inner LR is a local minimum}
We are now ready to extend the result of the previous section by showing that the inner LR must be a minimum of the effective potential if we restrict our attention to the geometries in which the outer LR is with a good approximation similar to the Kerr LR. In other words, rather than making any assumptions regarding the exotic physics that must be non-negligible at the inner LR scale, we impose constraints at the outer LR based on observational evidence. 

We require that the LR that mimics the Kerr LR is a saddle point of the effective potential that correspond to a minimum along the angular direction and a maximum along the radial direction. Failure to meet this property would have strong observational implications.

The property \eqref{eq:limit} of the effective potential implies the existence, for every $r$, of at least one local minimum $\bar{\theta}=\bar{\theta}(r)$ in the angular direction such that
\begin{equation}\label{eq:bartheta}
   \left. \frac{\partial}{\partial \theta} H_\pm(r,\theta)\right|_{\theta=\bar{\theta}}=0\,.
\end{equation}
This does not imply that $\bar{\theta}$ is part of a LR as generically the radial derivative will not vanish. The condition of the vanishing gradient of $H_\pm$ \eqref{eq:def_LR} is now 
\begin{equation}
       \frac{\partial}{\partial r} H_\pm(r,\bar{\theta})=0\,.
\end{equation}

\begin{figure*}

    \centering
    \includegraphics[width=7cm]{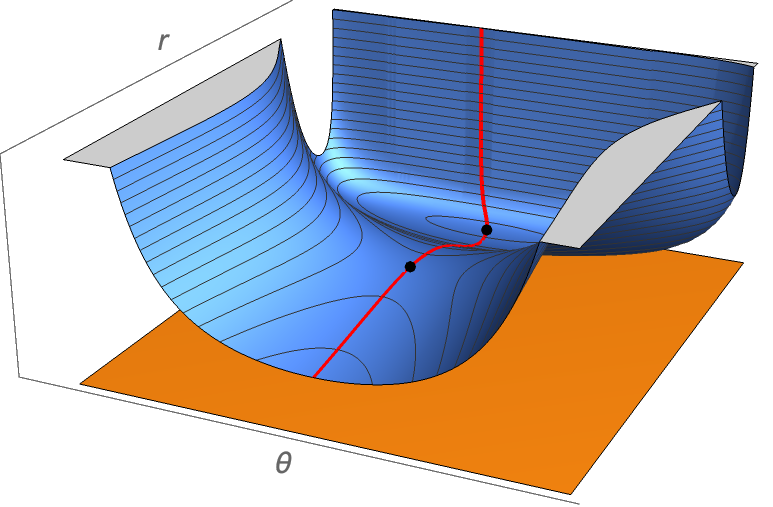}
    \caption{The figure shows the curve $\gamma$ (red line) for a specific choice of the potential  $H(r,\theta)=\frac{r^3+1}{10+(r/2)^4}\left(\frac{1}{\sin\theta
   }\right)^2$. This function has correct boundary behavior, but it is otherwise arbitrary and it is chosen for purely illustrative purposes. The LRs correspond to the two black dots. The outer one is a maximum along $\gamma$ and a saddle point in the 2-dimensional space $(r,\theta)$, while the inner one is a minimum both along gamma and for the 2-dimensional potential. }
    \label{fig:gamma}

\end{figure*}

We can define a curve $\gamma$ (see Fig.~\ref{fig:gamma}) that passes through the position of the outer LR and follows the locus of points in which Eq.~\eqref{eq:bartheta} is satisfied
\begin{equation}
    \gamma:r\mapsto \left(r,\bar{\theta}(r)\right)
\end{equation}
In the following we assume that $\gamma$ has no bifurcation point. If there are bifurcation points, we just need to repeat the argument for all the branches of the curve.

The directional derivative along $\gamma$, is the total derivative with respect to $r$, given by 
\begin{equation}
\begin{array}{cl}
\displaystyle    \frac{d}{dr}H_\pm\left(r,\bar{\theta}(r)\right)&=\displaystyle\partial_r H\left(r,\bar{\theta}(r)\right) + \partial_\theta H\left(r,\bar{\theta}(r)\right)\frac{d\bar{\theta}}{dr}\\
\\  &=\partial_r H\left((r,\bar{\theta}(r)\right)\,.
    \end{array}
\end{equation}
Therefore, the  LRs are at the points $(r,\bar{\theta})$ for which 
\begin{equation}
    \frac{d}{dr}H_{\pm}\left(r,\bar{\theta}(r)\right)=0\,.
\end{equation}
We can consider the same theorem used in the previous section, but now $v_\pm$ is a 1-dimensional map between the points on $\gamma$ and the 1-dimensional space spanned by $v_\pm=dH_\pm/dr$.
Therefore, the conservation of the Brouwer degree of this map implies
\begin{equation}\label{eq:Brower_gamma}
    \sum_n \frac{d}{dr}\frac{d}{dr}H_{\pm}\left(r_n,\bar{\theta}(r_n)\right)=0\,.
\end{equation}
This shows that there are as many maxima and minima along the direction given by $\gamma$. In general, some of these maxima and minima correspond to the extrema of the 2-dimensional effective potential, while others correspond to saddle points.

To study the stability of the LRs we now consider the second derivative along $\gamma$. A  very straightforward computation gives
\begin{equation}
\begin{array}{rl}
\displaystyle\hspace{-.2cm}\frac{d^2}{dr^2}H_{\pm}\left(r,\bar{\theta}(r)\right)=&\partial_r^2H_{\pm}\left(r,\bar{\theta}(r)\right)\\ \\ &\hspace{-2.55cm}+2\partial_r\partial_\theta H_{\pm}\left(r,\bar{\theta}(r)\right)\frac{d \bar{\theta}}{d r}
+\partial^2_\theta H_{\pm}\left(r,\bar{\theta}(r)\right)\left(\frac{d \bar{\theta}}{d r}\right)^2,
     \end{array}
\end{equation}
where we have omitted terms that vanish due to Eq.~\eqref{eq:bartheta}.
We now assume that this quantity is negative. This must be try the outer LR to mimic the LR of a black hole in general relativity. In fact, for a Kerr black hole $\bar{\theta}$ is independent of $r$. This can be explicitly checked, and it is a consequence of the fact that the geodesics equations are separable in the $r,\,\theta$ coordinates. Therefore, at the outer LR, where the geometry must mimic very closely Kerr spacetime,  we have  $d\bar{\theta}/{d r}\approx0$. Furthermore, for a Kerr black hole the LRs are saddle points of the effective potential and are maxima in the radial direction. Thus $\partial_r^2H_{\pm}\left(r,\bar{\theta}(r)\right)<0$, which imply
\begin{equation}\label{eq:maximum}
\frac{d^2}{dr^2}H_{\pm}\left(r_+,\bar{\theta}(r_+)\right)<0\,,
\end{equation}
where $r_+$ denotes the radius of the outer LR.
We assume that this inequality holds also for the UCO. Its violation would lead to observational effects.

Let us now consider the most natural case illustrated in Fig.~\ref{fig:gamma} in which the UCO possesses only 2 LRs. The conservation of the Brouwer degree implies that in total there is an equal number of minima and maxima along $\gamma$ and, by construction, the outer LR lies on gamma. Furthermore, Eq.~\eqref{eq:maximum} implies that the outer LR is a maximum along $\gamma$. So, the inner LR must be a minimum along gamma. 
We also know that from the point of view of the 2-dimension potential the outer LRs is a saddle point. Therefore the inner one must be an extremum and hence it is a minimum as it is a minimum along gamma.
Let us now show that even allowing for the presence of more than 2 LRs it is possible to show the presence of a stable LR, modulo adding an extra assumption on the potential at the outer LR scale.

Let us assume that, as is the case for the Kerr geometry, the outer LR is a global minimum in the $\theta$ direction for fixed $r$
\begin{equation}\label{Eq:cond_2}
    H_\pm\left(r_+,\bar{\theta}(r_+)\right)\leq H_\pm\left(r_+,\theta\right) \qquad\forall \,\,\theta\in(0,\pi)\,.
\end{equation}
With this assumption, the presence of a stable LR is almost trivial as the boundaries $r=0$, $\theta = \{0,\pi\}$ and $r=r_+$ create a potential barrier inside which there should be a minimum. We can find the position of the minimum following the curve $\gamma$ until the first local minimum along $\gamma$ (which must exist as proved previously) whose coordinate we can call $(r_-,\bar{\theta}(r_-))$. The value of the effective potential at that point is necessarily lower than its value at the outer LR as the derivative along $\gamma$ as we have moved from a local maximum to a local minimum (along gamma) without ever changing the sign of the derivative.  If this is a minimum of the effective potential we have reached the stable LR. Otherwise, it is a saddle point and there must be a direction along which  $H_\pm$ decreases. We can move along that direction and follow the path along which the gradient is negative and maximum in modulus. Given the potential is bounded from below, we have either to approach asymptotic infinity or a local minimum. The first possibility must be ruled out as we always moved in the direction of decreasing values of the potential. Therefore, starting from $(r_-,\bar{\theta}(r_-))$ it is not possible to reach values greater than $r_+$ as  
\begin{equation}
    H_\pm\left(r_-,\bar{\theta}(r_-)\right)<H_\pm\left(r_+,\bar{\theta}(r_+)\right)\leq H_\pm\left(r_+,\theta\right) \quad\forall \,\,\theta\,.
\end{equation}

This concludes the theorem and proves that there exists at least one stable LR.
No restriction regarding the dynamics or the matter content was necessary. Furthermore, we did not need to make any assumptions regarding the inner LR. Instead, we assumed that the UCOs mimic Kerr black holes at the outer LR scale. 
In particular, the simplest scenario in which there are only two LRs per rotation direction requires the property Eq.~\eqref{eq:maximum}. In case there are multiple LRs, we also need to assume the property \eqref{Eq:cond_2}.  
Crucially, we do not assume anything at the scale of the inner LR. 

\section{Discussion}
In this paper we have proved the following \\
\textit{\textbf{Theorem:}} \textit{If an axisymmetric UCO is invariant under simultaneous reflections $(t,\phi)\rightarrow(-t,-\phi)$ and it has LR that correspond to a minimum along the angular direction and a maximum along the radial direction such that Eqs.~\eqref{eq:maximum} and \eqref{Eq:cond_2} are satisfied, then there must exist at least one stable LR for each rotation sense.}

In practice, this result shows that axisymmetric objects compact enough to support a LR and that do not exhibit large deviations from general relativity at the outer LR scale must necessarily possess an additional stable LR. This reinforces the results of Ref.~\cite{Cunha:2017qtt} in which it was necessary to use the Einstein equations and the null energy condition to prove the stability of the inner LR. The proof is now completely geometrical and it applies to any modified theory of gravity irrespective of the matter content. 
\\
Of course, the null energy condition assumed in~\cite{Cunha:2017qtt} has a more clear physical interpretation than the geometrical assumptions in this manuscript. However, we do not need exotic matter to violate the null energy condition as that can be done by semiclassical gravity~\cite{Flanagan:1996gw}. Furthermore, assumptions at the inner LR can be impossible to test as current observations are not able to detect the presence of extra LRs. In converse,  all the assumptions of the theorem proved here are kinematical and at the scale of the outer LR. Therefore, while they have a less clear physical interpretation, these assumptions can be tested observationally. Therefore, this work does not supersede but rather complement the results of Ref.~\cite{Cunha:2017qtt} as both analysis have their strengths.

It is fascinating that it is possible to infer the eventual existence of an inner LR, which would be placed deep inside the strong gravity region where the properties of the geometry cannot yet be probed, with observations at the outer LR scale that can be tested with high precision. 

To get rid of the inner LR and the instability associated with it, one possibility would be to consider regular black holes rather than UCOs. However, regular black holes are unstable due to the presence of mass inflation instability \cite{PhysRevLett.63.1663,Brown:2011tv,Carballo-Rubio:2018pmi,Carballo-Rubio:2021bpr,DiFilippo:2022qkl,Carballo-Rubio:2024dca}. While it is possible to construct regular black holes geometries that are free from mass inflation and hence classically stable \cite{Carballo-Rubio:2022kad,Franzin:2022wai}, these geometries might suffer from semiclassical instabilities \cite{McMaken:2023uue,Balbinot:2023vcm}.

This does not necessarily imply that it is impossible to construct viable non-singular alternatives to black holes. First of all, there is a question regarding the timescales of the instabilities. While the instability at the inner horizon is expected to be very fast \cite{PhysRevLett.63.1663,Carballo-Rubio:2018pmi,Carballo-Rubio:2021bpr,DiFilippo:2022qkl,Carballo-Rubio:2024dca}, the LR instability is a non-linear process and so the timescale associated to the LR instability is model dependent and possibly long.

More exotic black holes mimickers, such as wormholes-like objects, are also possible~\cite{Simpson:2018tsi, Franzin:2021vnj,Mazza:2021rgq}. There are no known instabilities associated with these spacetimes. However, further studies are still needed to address the viability of such solutions and it is unclear how these objects can form as a result of gravitational collapse~\cite{Carballo-Rubio:2019fnb,Pandora,Carballo-Rubio:2021bpr}.

In conclusion, there are still numerous open issues in the quest for viable classes of black hole mimickers. This paper highlights and strongly reinforces one of them. While some may perceive this as problematic, it can alternatively be viewed as a promising aspect. With the inherent challenges of observational exploration deeper than the outer LR, lessons from theoretical analyses aimed at addressing consistency issues, integrated by lessons from observations, can effectively help us navigate the plethora of non-singular models, ultimately enriching our understanding of the nature of black holes.

\section{Acknowledgments}
I wish to express my gratitude to Luca Buoninfante, Raúl Carballo-Rubio, Pedro Cunha, Mohammad Ali Gorji, Carlos Herdeiro and Stefano Liberati for reading an early version of the draft and for providing useful feedback. 
I also acknowledge financial support from the PRIMUS/23/SCI/005 and UNCE24/SCI/016 grants by Charles University, and from the GAR 23-07457S grant from the Czech Science Foundation.

\bibliography{ref}
\end{document}